\documentclass[11pt,dvips,fleqn]{article}
\usepackage[dvips]{graphicx}
\usepackage{graphicx}
\usepackage{amssymb}

\date{\empty}

\title{{\normalsize\rightline{DESY 02-197}}
\vskip 1cm 
\bf Mini Black Holes from Ultrahigh Energy Cosmic Neutrinos\thanks{Talk 
presented at SUSY02, June 17 - 23, 2002, DESY Hamburg, Germany.}}
\author{Huitzu Tu \\
\vspace{4mm}  
\textsl{Deutsches Elektronen-Synchrotron DESY, Hamburg, Germany}}

\begin{document}
\maketitle

\vspace{-1cm}
\begin{abstract}

We review the perhaps most exciting phenomenology of models with extra
spatial dimensions and Planck scale near TeV: the production of mini
black holes in ultrahigh energy particle collisions, and the discovery
potential of cosmic ray/cosmic neutrino experiments for black hole events 
before the start of LHC. 
 
\end{abstract}

\section{Introduction}

It has been conjectured that mini black holes may be formed in 
particle collisions at energies higher than the Planck mass and with 
impact parameters smaller than a critical value
\cite{'tHooft:1987rb}.
In models with $\delta = D - 4$ extra spatial dimensions, where the 
Standard Model particles are assumed to reside on a 3-dimensional brane 
while only gravitons are allowed to propagate into the bulk, 
the Planck scale, which is the scale characterising quantum gravity,
can be just beyond the electroweak scale
\cite{Arkani-Hamed:1998rs,Randall:1999ee}.
Within such TeV-scale gravity models, above conjecture suggests that 
particle collisions at energies $\gtrsim$ TeV may result in the production 
of black holes of masses at this energy scale, provided the colliding 
particles come close enough
\cite{Argyres:1998qn}.

Due to their small masses, these microscopic black holes undergo decay 
processes rapidly. 
It is believed that these multi-dimensional black holes should 
Hawking-radiate 
\cite{Hawking:1974rv}
mainly into Standard Model particles on the brane rather than into the bulk
\cite{Emparan:2000rs}.
Thus direct observations of such black hole events are possible.
Estimates show that, depending on the value of the higher-dimensional
fundamental Planck scale, the Large Hadron Collider (LHC) may either turn 
into a black hole factory
\cite{Giddings:2002bu,Dimopoulos:2001hw}, 
where the black hole formation conjecture, the Hawking radiation law and 
the existence of extra spatial dimensions can be verified experimentally, 
or be able to put constraints on the model parameters from non-observation.
On the other hand, it is well known that particle astrophysics 
experiments are complementary to collider searches for new physics
beyond the Standard Model. 
In the case of black hole production in TeV-scale gravity models, one finds 
\cite{Feng:2002ib,Ringwald:2002vk,Anchordoqui:2001cg,Anchordoqui:2002ei,Kowalski:2002gb,Alvarez-Muniz:2002ga} 
that depending on the fluxes of the ultrahigh energy cosmic neutrinos, 
cosmic ray facilities such as Auger and neutrino telescopes like AMANDA 
and RICE may have an opportunity to see the first sign or put constraints on
black hole production parameters before LHC starts operating. 
IceCube has even discovery potential beyond the LHC reach.

In the following sections we give a brief review on the phenomenology of
black hole production and decay in the large extra dimension scenario 
\cite{Arkani-Hamed:1998rs}, and the prospects of the cosmic ray experiments
for detecting black hole events before LHC starts operating. 
More details and a more complete reference list can be found in e.g.
Ref.~\cite{Tu:2002xs}.

\section{Black hole production and decay in TeV-scale gravity}

TeV-scale gravity is a novel approach to the long-standing hierarchy 
problem.
The idea is to assume that the fundamental scale in physics is the TeV
scale, and there are $\delta \geq 1$ compact extra dimensions. 
The hierarchy between the four-dimensional Planck mass 
$M_{\rm pl} = (G_{\rm N} / \hbar)^{-1/2} \simeq 1.2 \cdot 10^{19}$ GeV
and the fundamental Planck scale $M_D \sim \textrm{TeV}$ 
arises either due to the large volume of the extra dimensions  
\cite{Arkani-Hamed:1998rs},
or through the ``warp factor'' arising from the background metric
\cite{Randall:1999ee}.

\subsection{Black hole production}

With the proposal of TeV-scale gravity, the remote possibility of probing 
the Planck scale physics is now within phenomenological reach.
In TeV-scale gravity models, the trans-Planckian energy regime corresponds 
to
\begin{equation}
   \sqrt{s} \gg M_D \hspace{1cm} \Rightarrow \hspace{1cm} 
   R_{\rm S} \gg \lambda_{\rm Pl} \gg \lambda_{\rm B}\, ,  
\end{equation}
where $\lambda_{\rm Pl}$ is the Planck length, $\lambda_{\rm B}$ the 
de Broglie wavelength, and 
\begin{equation}
   R_{\rm S} = \frac{1}{M_D} \left[\frac{\sqrt{s}}{M_D} \left(
   \frac{2^{\delta} \pi^{\frac{\delta - 3}{2}} \Gamma
   \left(\frac{3 + \delta}{2} \right)}{2 + \delta} \right)
   \right]^{\frac{1}{1 + \delta}}\,  
\label{Schwarzschild-radius} 
\end{equation}
is the Schwarzschild radius associated with the centre-of-mass (cm) energy 
$\sqrt{s}$ \cite{Myers:1986un}.
In this regime, gravitational interactions dominate over other gauge 
interactions. 
The gravitational scattering process in this regime is semiclassical and 
calculable by non-perturbative approaches only. 

The phenomenology of trans-Planckian energy scattering in large extra 
dimension scenarios has been studied in Ref.~\cite{Giudice:2001ce},
which focus on the regime of large impact parameter $b \gg R_S$, where the 
elastic cross section is calculable using the eikonal approximation.
On the other hand, in the regime where black hole formation is 
conjectured\footnote{String theory predicts that trans-Planckian energy 
scattering could lead to the creation of ``branes'' as well. 
For phenomenological investigations of $p$-brane production, see e.g.
\cite{Ahn:2002mj}.},
\begin{equation}
   \sqrt{s} \gg M_D\, , \hspace{0.4cm} b < R_{\rm S}\, ,
\end{equation}
exact calculations are impossible due to the high non-linearity of the 
Einstein equations.
Nevertheless, a geometrical parametrisation for the black hole production 
cross section at the parton-level $i j$,
\begin{equation}
  \sigma^{\rm bh}_{i j} (\hat{s}) 
  \approx \pi R^2_{\rm S} \left(M_{\rm bh} = \sqrt{\hat{s}} 
  \right)~\Theta \left(\sqrt{\hat{s}} - 
  M^{\rm min}_{\rm bh} \right)\, ,
\label{bhp_cs_parton}
\end{equation}
is believed to capture the essential features of this nonperturbative 
phenomenon
\cite{Eardley:2002re,Dimopoulos:2002qe}.
This semiclassical description is assumed to be valid above a minimum
black hole mass $M^{\rm min}_{\rm bh} \gg M_D$, which is taken to be a 
free parameter besides $M_D$ and $\delta = D - 4$.  
For the case $D = 4$, mass of the final state black hole is estimated 
to be $\sim {\cal O} (50 \% \div 80 \%)$ of the initial centre-of-mass 
energy $\sqrt{s}$
\cite{Eardley:2002re,Penrose:1974,D'Eath:1992hb}.
But estimate for the mass of the final black hole in $D > 4$ is not available
so far.
Besides, it is still not clear how to extend the study to the production 
of charged and/or spinning black holes in higher dimensions.

\subsection{Black hole decay}

Hawking has predicted that black holes should evaporate by thermally 
radiating real particles at the cost of their mass
\cite{Hawking:1974rv}.
For black holes produced with $M_{\rm bh} \gg M_D$ it is sufficient to 
adopt the semiclassical approximation for the purpose of estimating 
black hole event rates and event signatures in high-energy experiments,
since a black hole spends most of its lifetime in the stage where its 
mass is close to the initial value

Neglecting the backreaction of the emitted particles on the spacetime 
geometry (described by the greybody factor), a $(4 + \delta)$-dimensional 
Schwarzschild black hole of initial mass $M_{\rm bh} \gg M_D$ radiates 
thermally as a black body of surface area ${\cal A}_{\delta + 2}$ at the 
Hawking temperature $T_{\rm H}= (\delta + 1) / 4 \pi R_S$.
It is shown in Ref.~\cite{Emparan:2000rs}
that the multi-dimensional black holes localised on the brane radiate 
at equal rates
\begin{equation}
   \frac{d E}{d t} \simeq \sigma_{\delta + 4}\, 
   {\cal A}_{\delta + 2}\, T\,^{\delta + 4}_{\rm H} \propto 
   \frac{1}{R^2_{\rm S}}\, ,
\end{equation}
into a bulk field and into a brane field (the Stefan-Boltzmann constant in 
$(\delta + 4)$-dimensions $\sigma_{\delta + 4}$ is found to be almost 
independent of the number of extra dimensions).
The fact that there are much more fields on the brane than in the bulk then
leads to the conclusion that small black holes localised on the brane  
radiate mainly into Standard Model particles on the brane rather than 
into the bulk. 
Approximately $\left< n \right> \approx \frac{M_{\rm bh}}{2~T_{\rm H}}$ 
particles \cite{Dimopoulos:2001hw}, mostly hadrons and leptons, will be 
emitted during $\tau_D \sim 10^{-26}$ s, the lifetime of an average 
mini black hole.

\section{Mini black holes at colliders and from cosmic neutrinos}

If the fundamental Planck scale $M_D$ is below 2 TeV,
the Large Hadron Collider (LHC), with its design values 
$\sqrt{s} = 14$ TeV for the proton-proton cm energy 
and ${\cal L} = 10^{34}~{\rm cm}^{-2}~{\rm s}^{-1}$ for the luminosity,
will be producing mini black holes copiously.
The unique signatures of black hole decay (highly isotropical events, with
characteristic spectra and species ratios)
\cite{Giddings:2002bu,Dimopoulos:2001hw}
should then enable the discrimination against backgrounds from any known 
extension of the Standard Model.  

\begin{figure}[ht!]
\vspace{-1.3cm}
\includegraphics[width=7.95cm]{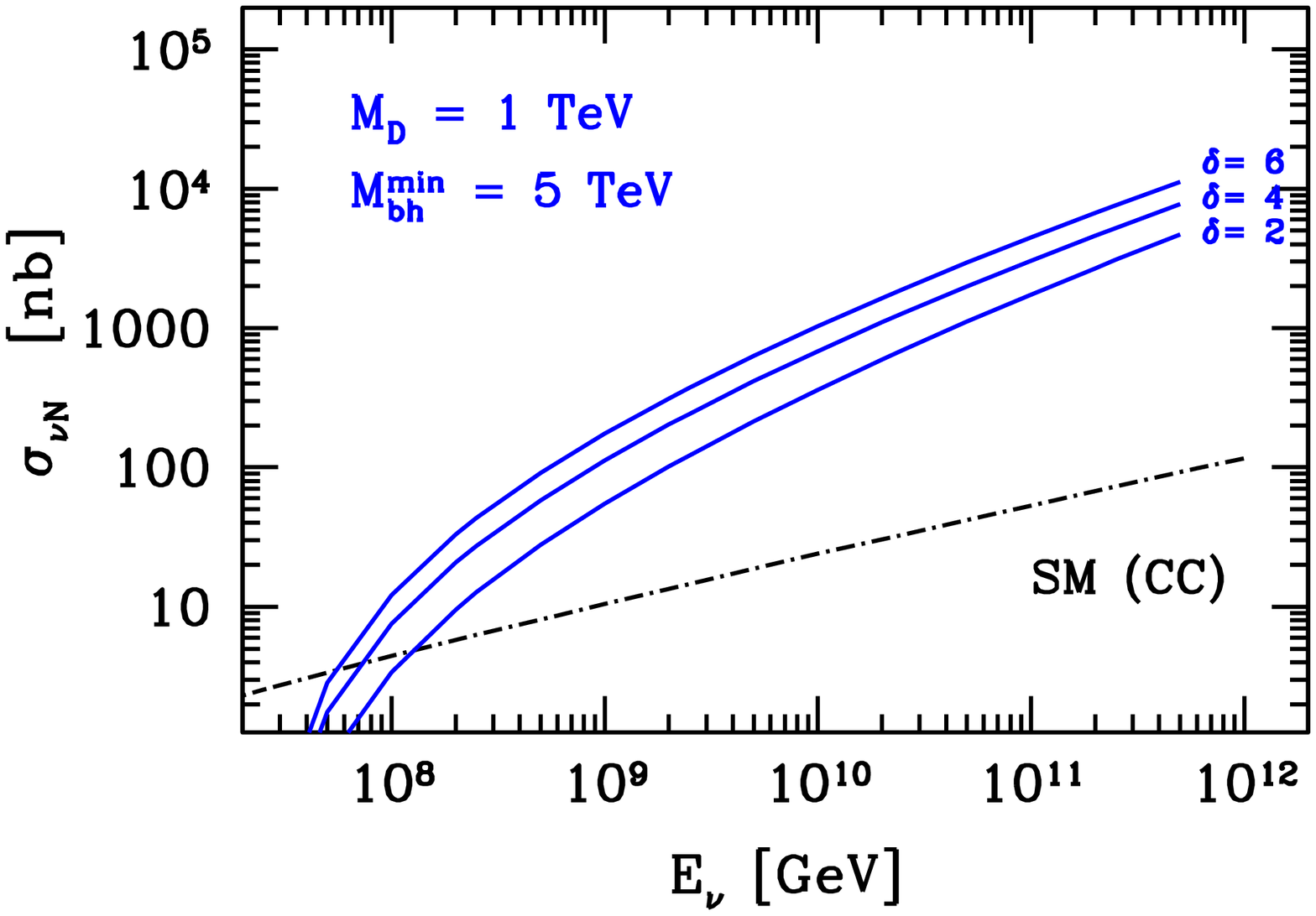}
\includegraphics[width=7.95cm]{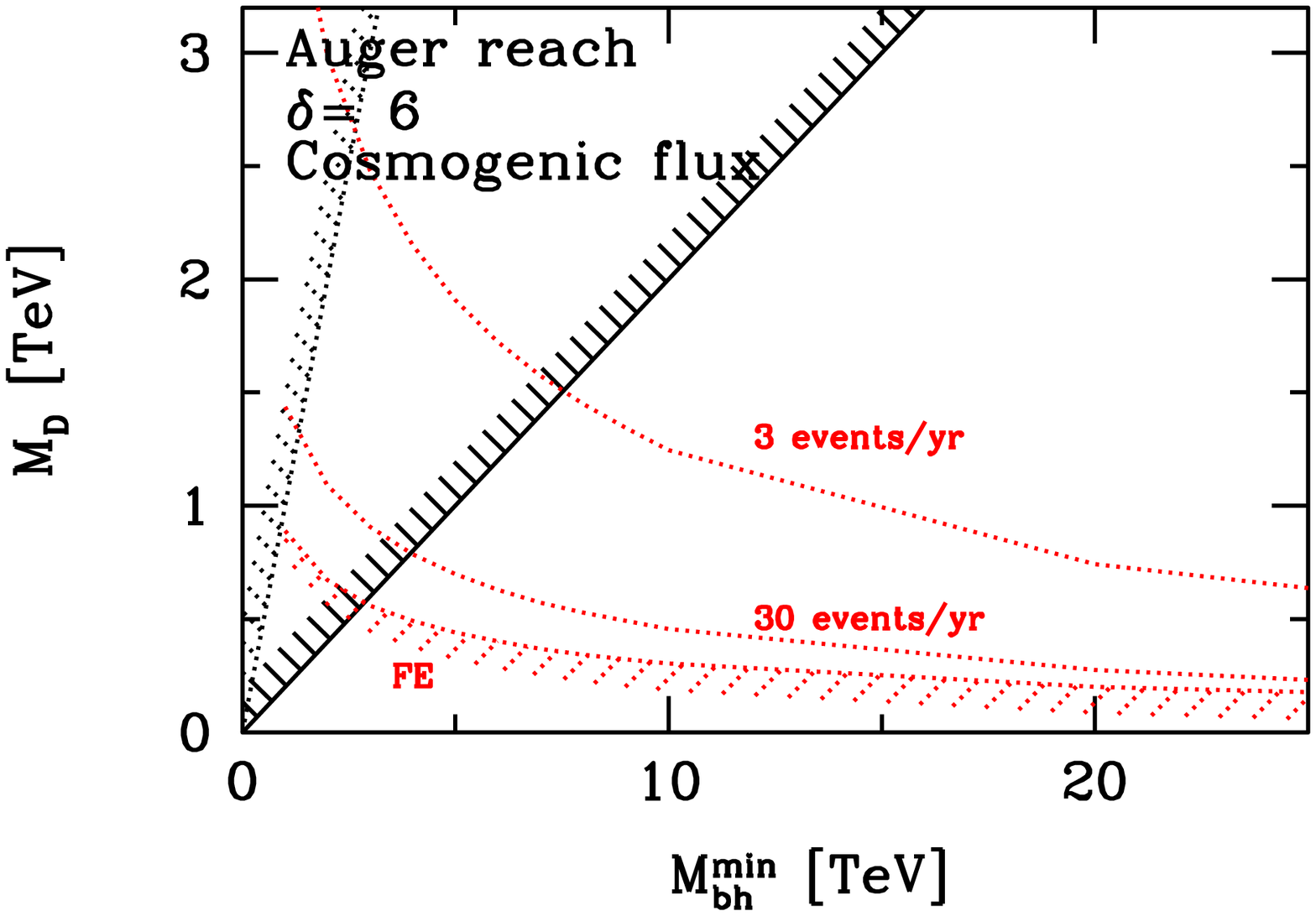}
\vspace{-1.3cm}
\caption[dum]{\small
Left: Cross section for black hole production in neutrino-nucleon 
scattering, as a function of the incident neutrino energy. 
Right: Projected Auger reach in the black hole production parameters for
$\delta = 6$ large extra dimensions, by exploiting the cosmogenic neutrino 
flux from Ref.~\protect\cite{Protheroe:1996ft} with cutoff energy 
$3 \times 10^{21}$ eV for the ultrahigh energy cosmic rays.
The shaded dotted, $M_D = M^{\rm min}_{\rm bh}$, and shaded solid,
$M_D = (1 / 5) M^{\rm min}_{\rm bh}$, lines give a rough indication of the
boundary of applicability of the semiclassical picture  
\cite{Giddings:2002bu}.
Also shown is the constraint arising from the non-observation of horiontal
air showers by the Fly's Eye collaboration (shaded dotted line labeled 
``FE'').
The constraint imposed by AGASA obtained in 
Ref.~\protect\cite{Anchordoqui:2001cg}
lies slightly above the 30 events/yr contour line for Auger.
}
\label{bh_cs_auger}
\end{figure}

However, until the LHC starts operating, cosmic rays provide the only 
access to the required energy scales.
Cosmic rays of energies up to $\simeq 10^{21}$ eV have been observed.
The ``cosmogenic'' neutrinos, expected from the cosmic ray interactions 
with the Cosmic Microwave Background (e.g. $p \gamma \rightarrow \Delta 
\rightarrow n \pi^+ \rightarrow \nu_{\mu} \bar{\nu}_{\mu} \nu_e...$), 
are more or less guaranteed to exist among ultrahigh energy cosmic 
neutrinos predicted from various sources (for recent reviews, see 
Ref.~\cite{Protheroe:1999ei}).
Thus, if TeV-scale gravity is realised, ultrahigh energy cosmic rays/cosmic 
neutrinos should have been producing mini black holes in the atmosphere 
throughout earth's history. 
For cosmic ray facilities such as Fly's Eye, AGASA and Auger, the clearest 
black hole signals are neutrino-induced quasi-horizontal air showers which 
occur at rates exceeding the Standard Model rate by a factor of $10 - 10^2$ 
(see Fig.~\ref{bh_cs_auger} (left)), and have distinct characteristics 
\cite{Feng:2002ib,Ringwald:2002vk,Anchordoqui:2001cg,Anchordoqui:2002ei}. 
Black hole production could also enhance the detection rate at neutrino 
telescopes such as AMANDA/IceCube, ANTARES, Baikal and RICE significantly, 
both of contained and of through-going events
\cite{Kowalski:2002gb,Alvarez-Muniz:2002ga}. 

The reach of cosmic ray facilites in the black hole production has been 
investigated in detail \cite{Ringwald:2002vk,Anchordoqui:2001cg} by
exploiting the cosmogenic neutrino fluxes.
It is argued in Ref.~\cite{Anchordoqui:2001cg} that an excess of a handful
of quasi-horizontal events are sufficient for a discrimination against the
Standard Model background.
An inspection of Fig.~\ref{bh_cs_auger} (right) thus leads to the conclusion
that, already for an ultrahigh energy neutrino flux at the cosmogenic level 
estimated in Ref.~\cite{Protheroe:1996ft},
the Pierre Auger Observatory, expected to become fully operational in 
2003, has the opportunity to see first signs of black hole production. 

On the other hand, the non-observation of horizontal air showers reported 
by the Fly's Eye and the AGASA collaboration provides a stringent bound on 
$M_D$, which is competitive with existing bounds on $M_D$ from colliders 
as well as from astrophysical and cosmological considerations, particularly 
for larger numbers of extra dimensions ($\delta \geq 5$) and smaller 
threshold ($M^{\rm min}_{\rm bh} \lesssim 10$ TeV) for the semiclassical 
description, eq.~(\ref{bhp_cs_parton}).

\begin{figure}[ht!]
\vspace{-1.3cm}
\includegraphics[width=7.95cm]{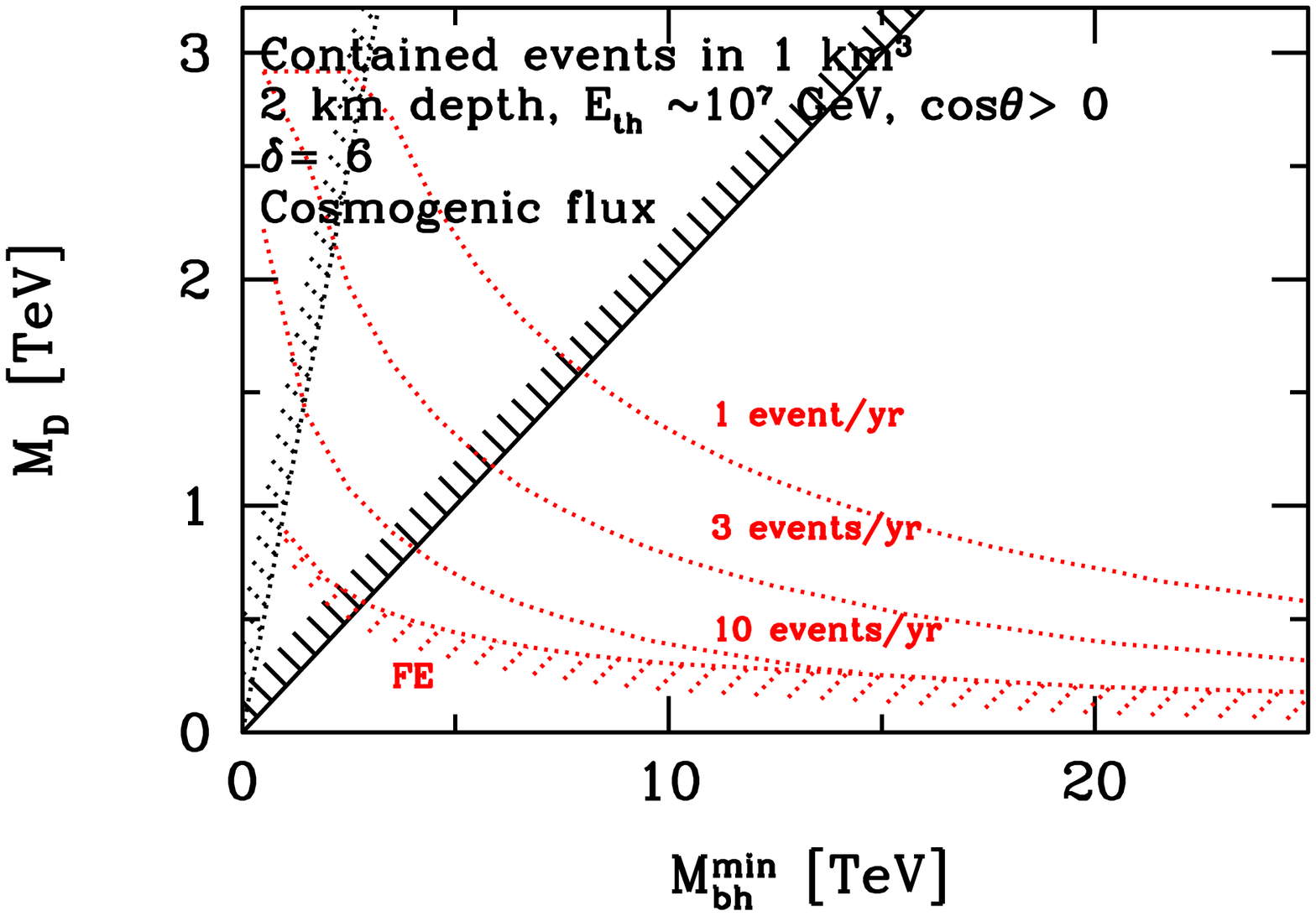}
\includegraphics[width=7.95cm]{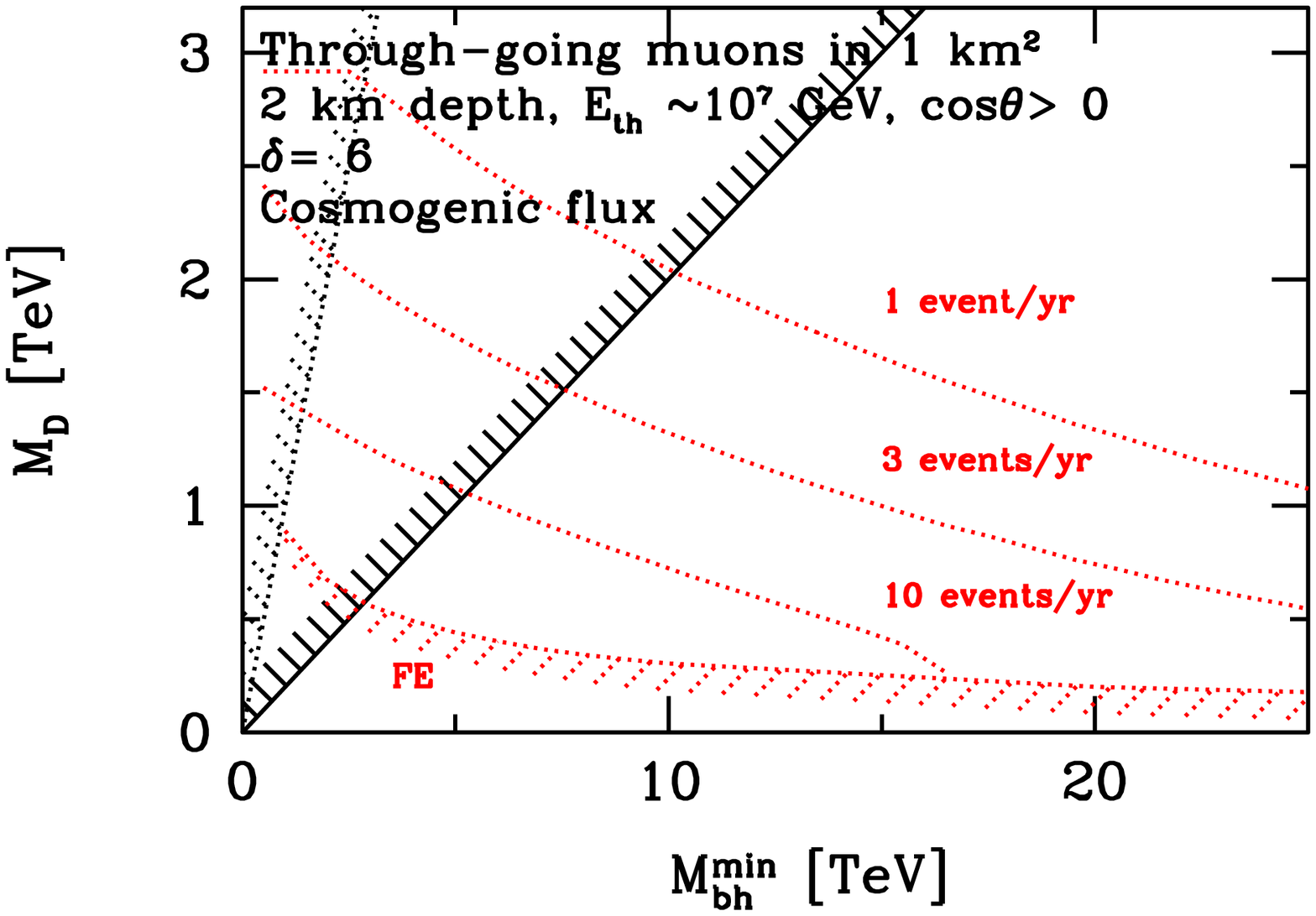}
\vspace{-1.3cm}
\caption[dum]{\small
Reach of the neutrino telescopes in the black hole production parameters 
for $\delta = 6$ large extra dimensions, with the shaded dotted, 
$M_D = M^{\rm min}_{\rm bh}$, shaded solid, $M_D = (1 / 5) 
M^{\rm min}_{\rm bh}$, lines and the shaded dotted line labeled ``FE'' 
same as in Fig.~\ref{bh_cs_auger} (right).
Left: for contained events in an under-ice detector at a depth of 2 km and 
with an 1 km$^3$ fiducial volume.
Right: for through-going muons in an under-ice detector at a depth of 2 km 
and with an 1 km$^2$ effective area.
Both by exploiting the cosmogenic neutrino flux from 
Ref.~\cite{Protheroe:1996ft} with cutoff energy $3 \times 10^{21}$ eV 
for the ultrahigh energy cosmic rays.
}
\label{bh_par_nu_tel}
\end{figure}

As for neutrino telescopes, investigations (see Fig.~\ref{bh_par_nu_tel}
(left)) show that due to their small volume $V \approx 0.001 \div 0.01$ km$^3$ 
for contained events, the currently operating underwater/-ice neutrino 
telescopes AMANDA and Baikal cannot yield a large enough contained event
rate to challenge the already existing limits from Fly's Eye and AGASA.
Even IceCube does not seem to be really competitive, since the final 
effective volume $V \approx 1$ km$^3$ will be reached only after the LHC
starts operating and Auger has taken data for already a few years.
But sensible constraints on black hole production can be expected from RICE, 
a currently operating radio-Cherenkov neutrino detector with an effective 
volume $\approx 1$ km$^3$ for $10^8$ GeV electromagnetic cascades, using 
already availabe data. 

The ability to detect muons from distant neutrino reactions increases an
underwater/-ice detector's effective neutrino target volume dramatically.
In the case that the neutrino flux is at the level of the cosmogenic one,
only a few $(\lesssim 1)$ events from Standard Model background are
expected per year.
Thus, with an effective area of about 0.3 km$^2$ for down-going muons above
$10^7$ GeV and 5 years data available, AMANDA should be able to impose 
strong constraints if no through-going muons above $10^7$ GeV are seen in
the currently available data (see Fig.~\ref{bh_par_nu_tel} (right)).
Moreover, in the optimistic case that an ultrahigh energy cosmic neutrino 
flux significantly higher than the cosmogenic one is realised in nature,
one even has discovery potential for black holes at IceCube beyond the 
reach of LHC, though discrimination between Standard Model background
and black hole events becomes crucial.

\section{Conclusion}

TeV-scale gravity models offer the first opportunity to test the conjecture 
of black hole formation in trans-Planckian energy collisions and the
prediction of Hawking radiation at colliders.
The LHC will be producing black holes copiously if the fundamental Planck 
scale $M_D$ is below 2 TeV, while the reach of the cosmic ray facilities 
and the neutrino telescopes depends on the unknown ultrahigh energy cosmic
neutrino fluxes.
It is found that, already for an ultrahigh energy neutrino flux at the 
level expected from cosmic ray interactions with the cosmic microwave 
background radiation, cosmic ray experiments are able to put sensible 
constraints on black hole production parameters and/or bounds on 
TeV-scale gravity, which are among the most stringent ones to date.


\begin{thebibliography}{99}

\bibitem{'tHooft:1987rb}
G.~'t Hooft,
Phys.\ Lett.\ B {\bf 198} (1987) 61;
%
Nucl.\ Phys.\ B {\bf 304} (1988) 867;
%
D.~Amati, M.~Ciafaloni and G.~Veneziano,
Phys.\ Lett.\ B {\bf 197} (1987) 81;
%
Int.\ J.\ Mod.\ Phys.\ A {\bf 3} (1988) 1615;
%
Phys.\ Lett.\ B {\bf 216} (1989) 41;
%
Nucl.\ Phys.\ B {\bf 347} (1990) 550;
%
Nucl.\ Phys.\ B {403} (1993) 707.
%

\bibitem{Arkani-Hamed:1998rs}
N.~Arkani-Hamed, S.~Dimopoulos and G.~R.~Dvali,
Phys.\ Lett.\ B {\bf 429} (1998) 263
[arXiv:hep-ph/9803315];
%
I.~Antoniadis, N.~Arkani-Hamed, S.~Dimopoulos and G.~R.~Dvali,
Phys.\ Lett.\ B {\bf 436} (1998) 257
[arXiv:hep-ph/9804398].
%

\bibitem{Randall:1999ee}
L.~Randall and R.~Sundrum,
Phys.\ Rev.\ Lett.\  {\bf 83} (1999) 3370
[arXiv:hep-ph/9905221].
%


\bibitem{Argyres:1998qn}
P.~C.~Argyres, S.~Dimopoulos and J.~March-Russell,
Phys.\ Lett.\ B {\bf 441} (1998) 96
[arXiv:hep-th/9808138];
%
T.~Banks and W.~Fischler,
arXiv:hep-th/9906038.
%

\bibitem{Hawking:1974rv}
S.~W.~Hawking,
Nature {\bf 248} (1974) 30;
%
Commun.\ Math.\ Phys.\  {\bf 43} (1975) 199.
%

\bibitem{Emparan:2000rs}
R.~Emparan, G.~T.~Horowitz and R.~C.~Myers,
Phys.\ Rev.\ Lett.\  {\bf 85} (2000) 499
[arXiv:hep-th/0003118].
%

\bibitem{Giddings:2002bu}
S.~B.~Giddings and S.~Thomas,
Phys.\ Rev.\ D {\bf 65} (2002) 056010
[arXiv:hep-ph/0106219].
%

\bibitem{Dimopoulos:2001hw}
S.~Dimopoulos and G.~Landsberg,
Phys.\ Rev.\ Lett.\  {\bf 87} (2001) 161602
[arXiv:hep-ph/0106295].
%

\bibitem{Feng:2002ib}
J.~L.~Feng and A.~D.~Shapere,
Phys.\ Rev.\ Lett.\  {\bf 88} (2002) 021303
[arXiv:hep-ph/0109106].
%

\bibitem{Ringwald:2002vk}
A.~Ringwald and H.~Tu,
Phys.\ Lett.\ B {\bf 525} (2002) 135
[arXiv:hep-ph/0111042].
%

\bibitem{Anchordoqui:2001cg}
L.~A.~Anchordoqui, J.~L.~Feng, H.~Goldberg and A.~D.~Shapere,
Phys.\ Rev.\ D {\bf 65} (2002) 124027
[arXiv:hep-ph/0112247];
%
arXiv:hep-ph/0207139.
%


\bibitem{Anchordoqui:2002ei}
L.~Anchordoqui and H.~Goldberg,
Phys.\ Rev.\ D {\bf 65} (2002) 047502
[arXiv:hep-ph/0109242].
%


\bibitem{Kowalski:2002gb}
M.~Kowalski, A.~Ringwald and H.~Tu,
Phys.\ Lett.\ B {\bf 529} (2002) 1
[arXiv:hep-ph/0201139].
%

\bibitem{Alvarez-Muniz:2002ga}
J.~Alvarez-Muniz, J.~L.~Feng, F.~Halzen, T.~Han and D.~Hooper,
Phys.\ Rev.\ D {\bf 65} (2002) 124015
[arXiv:hep-ph/0202081];
%
S.~I.~Dutta, M.~H.~Reno and I.~Sarcevic,
Phys.\ Rev.\ D {\bf 66} (2002) 033002
[arXiv:hep-ph/0204218].
%

\bibitem{Tu:2002xs}
H.~Tu,
to appear in the Proceedings of the 5th Moscow International School of Physics
and 30th ITEP Winter School of Physics, Moscow, Russia, 2002, 
arXiv:hep-ph/0205024.
%

\bibitem{Myers:1986un}
R.~C.~Myers and M.~J.~Perry,
Annals Phys.\  {\bf 172} (1986) 304.
%

\bibitem{Giudice:2001ce}
G.~F.~Giudice, R.~Rattazzi and J.~D.~Wells,
Nucl.\ Phys.\ B {\bf 630} (2002) 293
[arXiv:hep-ph/0112161];
%
R.~Emparan, M.~Masip and R.~Rattazzi,
Phys.\ Rev.\ D {\bf 65} (2002) 064023
[arXiv:hep-ph/0109287].
%

\bibitem{Ahn:2002mj}
E.~J.~Ahn, M.~Cavaglia and A.~V.~Olinto,
arXiv:hep-th/0201042;
%
P.~Jain, S.~Kar, S.~Panda and J.~P.~Ralston,
arXiv:hep-ph/0201232;
%
L.~A.~Anchordoqui, J.~L.~Feng and H.~Goldberg,
Phys.\ Lett.\ B {\bf 535} (2002) 302
[arXiv:hep-ph/0202124];
%
K.~Cheung,
Phys.\ Rev.\ D {\bf 66} (2002) 036007
[arXiv:hep-ph/0205033].
%






\bibitem{Eardley:2002re}
D.~M.~Eardley and S.~B.~Giddings,
Phys.\ Rev.\ D {\bf 66} (2002) 044011
[arXiv:gr-qc/0201034].
%

\bibitem{Dimopoulos:2002qe}
S.~Dimopoulos and R.~Emparan,
Phys.\ Lett.\ B {\bf 526} (2002) 393
[arXiv:hep-ph/0108060].
%

\bibitem{Penrose:1974}
R.~Penrose, presented at the Cambridge University Seminar,
Cambridge, England, 1974 (unpublished).
%


\bibitem{D'Eath:1992hb}
P.~D.~D'Eath and P.~N.~Payne,
Phys.\ Rev.\ D {\bf 46} (1992) 658;
%
Phys.\ Rev.\ D {\bf 46} (1992) 675;
%
Phys.\ Rev.\ D {\bf 46} (1992) 694.
%


\bibitem{Protheroe:1999ei}
R.~J.~Protheroe,
Nucl.\ Phys.\ Proc.\ Suppl.\  {77} (1999) 465;
%
R.~Gandhi,
Nucl.\ Phys.\ Proc.\ Suppl.\  {91} (2000) 453;
%
J.~G.~Learned and K.~Mannheim,
Ann.\ Rev.\ Nucl.\ Part.\ Sci.\  {50} (2000) 679.
%

\bibitem{Protheroe:1996ft}
R.~J.~Protheroe and P.~A.~Johnson,
Astropart.\ Phys.\  {4} (1996) 253
[Erratum-ibid.\  {5} (1996) 215].
%


\end{thebibliography}
\end{document}